\documentclass[prb,letterpaper,aps,superscriptaddress,floatfix,twocolumn,nofootinbib]{revtex4-2}
\pdfoutput=1
\usepackage{graphicx}
\usepackage{amsmath}
\usepackage{amsfonts}
\usepackage{amssymb}
\usepackage[normalem]{ulem}
\usepackage[small,bf]{subfigure}
\usepackage[utf8]{inputenc}
\usepackage{dsfont}
\usepackage{xcolor}
\usepackage{slashed}
\usepackage{soul}
\usepackage{comment}
\usepackage{tikz-cd}
\usepackage{hyperref}

\newcommand{\beq}{\begin{equation}}
\newcommand{\eeq}{\end{equation}}

\newcommand{\bR}{{{\bf{R}}}}

\newcommand{\beqa}{\begin{eqnarray}}
\newcommand{\eeqa}{\end{eqnarray}}

\newcommand{\ea}{\end{array}}

\def\eea{\end{eqnarray}}

\def\<{\langle}
\def\>{\rangle}

\def\bR{\mathbb{R}}

\usepackage{amsmath}
\usepackage{comment}
\usepackage{amssymb}
\usepackage{amsthm}

\theoremstyle{definition}

\usepackage{color}
\usepackage{tikz-cd}
\usepackage{comment}

\usepackage{tikz}
\usepackage[export]{adjustbox}

\def\[#1\]{%
  \begin{equation}\begin{gathered}#1\end{gathered}\end{equation}%
}

\usepackage[safe]{tipa}
\allowdisplaybreaks

\begin{document}
\title{Exact Chiral Symmetries of 3+1D Hamiltonian Lattice Fermions}
\author{Lei Gioia}
\affiliation{Walter Burke Institute for Theoretical Physics, Caltech, Pasadena, CA, USA}
\affiliation{Department of Physics, Caltech, Pasadena, CA, USA}
\author{Ryan Thorngren}
\affiliation{Mani L. Bhaumik Institute for Theoretical Physics, Department of Physics and Astronomy,
University of California, Los Angeles, CA 90095, USA}

\begin{abstract}
We construct Hamiltonian models on a 3+1D cubic lattice for a single Weyl fermion and for a single Weyl doublet protected by exact (as opposed to emergent) chiral symmetries. In the former, we find a not-on-site, non-compact chiral symmetry which can be viewed as a Hamiltonian analog of the Ginsparg-Wilson symmetry in Euclidean lattice models of Weyl fermions. In the latter, we combine an on-site $U(1)$ symmetry with a not-on-site $U(1)$ symmetry, which together generate the $SU(2)$ flavor symmetry of the doublet at low energies, while in the UV they generate an algebra known in integrability as the Onsager algebra. This latter model is in fact the celebrated magnetic Weyl semimetal which is known to have a chiral anomaly from the action of $U(1)$ and crystalline translation, that gives rise to an anomalous Hall response - however reinterpreted in our language, it has two exact $U(1)$ symmetries that gives rise to the global $SU(2)$ anomaly which protects the gaplessness even when crystalline translations are broken. We also construct an exact symmetry-protected single Dirac cone in 2+1D with the $U(1) \rtimes T$ parity anomaly. Our constructions evade both old and recently-proven no-go theorems by using not-on-siteness in a crucial way, showing our results are sharp.

\end{abstract}

\maketitle

\section{Introduction}

Regulating chiral gauge theories like the Standard Model on the lattice has been a long standing problem. As well as offering a route for extracting more precise predictions from these theories, {it also highlights} a deep theoretical problem thanks to early no-go results like the Nielsen-Ninomiya theorem \cite{karsten1981lattice,nielsen1981absence,NIELSEN1981173}.

{ This theorem in its basic form says that in any lattice system of fermions with an on-site $U(1)$ symmetry, that symmetry must act non-chirally in the IR. In particular, if the model has only charge 1 Weyl fermions in the IR, then they must come in equal numbers of left-handed and right-handed chiralities. This is known as \textit{fermion doubling}.

For quadratic/non-interacting/free Hamiltonian models, we can give an intuitive reason for this phenomenon: Weyl fermions appear in the Brillouin zone as monopoles for the Berry connection \cite{Armitage_2018,berry1984quantal}. Their total number must therefore be zero because the Brillouin zone is compact and the Berry magnetic field lines form closed loops, just like ordinary magnetic fields. 

Fermion doubling remains a problem even if we want to construct chiral gauge theories with vanishing gauge anomalies, such as the Standard Model, so long as we insist on using free fermions. See Appendix \ref{appfermiondoubling}.

}

There have been several approaches so far to circumvent the fermion doubling problem. A simple strategy is to try to ``gap out'' the unwanted fermions by introducing carefully chosen mass terms. Examples of this construction are Wilson fermions \cite{wilsonfermions} and Kogut-Susskind fermions \cite{KogutSusskind}.

{ As a consequence of the Nielsen-Ninomiya theorem, these mass terms necessarily break any on-site symmetries which would otherwise become chiral symmetries.} This puts the fermions in danger of being \emph{fine tuned}, unless there is enough remnant symmetry to protect them from obtaining other mass terms (i.e. to prevent additive mass renormalization). { For example there could be} discrete crystalline symmetries of the lattice model such as translation, which can act chirally. This mechanism stabilizes gapless fermions in condensed matter systems such as Weyl semimetals \cite{RevModPhys.90.015001,Burkov_ARCMP,zhang2016signatures,PhysRevLett.124.096603,PhysRevResearch.3.043067}.

Another method to stabilize such fermions is to use more general \emph{not-on-site} symmetries, which can give rise to continuous chiral symmetries. An early example is the Ginsparg-Wilson symmetry of a Euclidean lattice fermion \cite{ginsparg1982remnant,luscher1998exact} which as { an infinitesimal transformation} takes the form
\[\label{eqn-ginsparg-wilson-symmetry}\delta \psi = \gamma^5 (1 - \frac12 aD)\psi,\]
where $a$ is the lattice spacing, and $D$ is a finite difference operator which is a discretization of the spacetime Dirac operator. The operator $(D\psi)(x)$ is a (linear) function of $\psi$ at nearby spacetime lattice points, which makes it not-on-site, whereas an on-site symmetry would be a function of $\psi(x)$ alone, such as the vector symmetry. When $a \to 0$ this becomes the continuum axial symmetry of a Dirac fermion. It thus protects the Dirac fermion from gaining a mass. To emphasize that this axial symmetry is not just emergent, but that it is forced by the UV Ginsparg-Wilson symmetry, we call it an ``emanant symmetry'' after \cite{seibergemanant}.

In this paper, we will construct Hamiltonian models with minimal number of fermions at low energy, transforming anomalously under not-on-site symmetries. In particular, we construct two main models:
\begin{enumerate}
    \item An ultralocal\footnote{Ultralocal means finite range in real space.} 3+1D model with a single Weyl fermion in the IR, protected by an emanant $U(1)$ chiral symmetry which is also ultralocally generated but not quantized in the UV
    \item An ultralocal 3+1D model with a pair of Weyl fermions, protected by an emanant $SU(2)$ chiral symmetry with Witten's global $SU(2)$ anomaly \cite{witten19822}. The $SU(2)$ has a pair of ultralocal generators which are quantized but which do not close to form $su(2)$, instead generating an infinite-dimensional Lie algebra known as the Onsager algebra.
    
\end{enumerate}
We emphasize that these models have the minimal number of fermions for their emanant anomalous symmetries, and that each is protected from gaining a mass---they are not fine tuned. In the Appendix, we consider models with time reversal symmetry in 2+1D and 3+1D, but have to relax ultralocality.

Other Hamiltonian families of models of fermions enjoying Ginsparg-Wilson symmetries exist, such as overlap fermions \cite{neuberger1998exactly,kaplan2012chiralsymmetrylatticefermions,clancy2024toward,creutz2001newfermionhamiltonianlattice,singh2025ginspargwilsonhamiltoniansimprovedchiral}. { However, these models are not ultralocal, and can be difficult to compute with. As far as we know, our constructions 1 and 2 give the first ultralocal models with these minimal representations and emanant continuous symmetry groups. 

Furthermore, our methods are rather simple, and open a new approach to constructing ultralocal Hamiltonian lattice models with Ginsparg-Wilson-like symmetries.} From the condensed matter point of view, one can interpet our models as Weyl semimetals~\cite{RevModPhys.90.015001,Burkov_ARCMP}. Using band theory, and especially the Bogoliubov-de-Gennes (BdG) formalism, we are able to study a large class of transformations generalizing the Ginsparg-Wilson symmetry \eqref{eqn-ginsparg-wilson-symmetry} which are linear in the fermion operators, which we leverage to produce the models. We can also derive some no-go results in this formalism, constraining what these not-on-site symmetries must look like for certain anomalous symmetries to emanate from them, making contact with other no-go theorems generalizing Nielsen-Ninomiya such as in \cite{Fidkowski_2023}.

\section{Symmetry-protected single Weyl fermion in 3+1D}
\label{sec:symprot1weyl}

{
Here we build a (time-reversal broken) 3+1D tight-binding model with finite-range hopping and a single Weyl node at crystalline momentum $\mathbf{k}=\mathbf{0}$ that is protected by a finite-range non-on-site chiral symmetry. The Hamiltonian and symmetry we will construct both lack the full cubic symmetry of the lattice and only retain the $\frac{\pi}{2}$ rotational symmetry around the $\hat{z}$ axis. In Appendix~\ref{app:cubicsym} we present a single symmetry-protected Weyl fermion with the full cubic symmetry. However, we first present the model below because it is much simpler and will inspire the construction of Section \ref{secSU2}.
}

We start with a two-band model (two fermion species per site, which we think of as spin $s\in\{\uparrow,\downarrow\}$) on a cubic lattice, known as a magnetic Weyl semimetal, described by the second-quantized Hamiltonian
\begin{align}
    H_2=\sum_{\mathbf{k},s,s'}c^\dagger_{\mathbf{k},s} h_2(\mathbf{k})_{ss'} c_{\mathbf{k},s'}\quad,
\end{align}
where $h_2(\mathbf{k})$, the Bloch Hamiltonian, is given by
\begin{align}\label{eqnh2}
    h_2(\mathbf{k})= &\sin k_x\sigma^x+\sin k_y \sigma^z +\left[\sin k_z+m(\mathbf{k})\right]\sigma^y\quad,
\end{align}
with $m(\mathbf{k})={{\Delta}}(2-\cos k_x-\cos k_y)$ {for $\Delta\in \mathbb{R}$}, and $\boldsymbol{\sigma}=(\sigma^x,\sigma^y,\sigma^z)$ are Pauli matrices acting on the spin degree of freedom. It hosts two Weyl fermions at low energy { when $|\Delta|> 1$}, seen by linearizing the Hamiltonian are the node at momentum $\mathbf{k}_1\equiv\mathbf{0}$ and another at momentum $\mathbf{k}_2\equiv(0,0,\pi)$.

This model has an on-site $U(1)$ symmetry $c_{\mathbf{r}} \mapsto e^{-i\theta \hat{Q}_0} c_{\mathbf{r}} e^{i\theta \hat{Q}_0}=e^{i\theta} c_{\mathbf{r}}$, generated by the on-site charge $\hat{Q_0}=\sum_{\mathbf{r},s}c_{\mathbf{r},s}^\dag c_{\mathbf{k},s}$. We will need to break this in order to gap the Weyl node at $\mathbf{k}_2$. In order to do this, let us write our Hamiltonian in the BdG formalism with the basis $d^\dag_\mathbf{k}\equiv(c^\dag_{\mathbf{k}\uparrow},c^\dag_{\mathbf{k}\downarrow},c_{-\mathbf{k}\uparrow},c_{-\mathbf{k}\downarrow})$, such that the Hamiltonian takes the form
\begin{align}
    h_2^{\mathrm{BdG}}(\mathbf{k})= &\frac{1}{2}\left[\sin k_x\sigma^x+\sin k_y \sigma^z+\sin k_z\tau^z\sigma^y +m(\mathbf{k})\sigma^y\right].
\end{align}
The Pauli matrices $\tau^{x,y,z}$ act on a fictitious doubling degree of freedom which separately labels particles at $\textbf{k}$ and holes at $-\textbf{k}$ (this is also where the factor of $\frac12$ comes from). Thus, any valid Hamiltonian \emph{or symmetry generator} $h^{\mathrm{BdG}}(\mathbf{k})$ in this BdG formalism must satisfy a particle-hole {redundancy}
\begin{align}
    \tau^x h^{\mathrm{BdG}}(\mathbf{k})^T \tau^x=-h^{\mathrm{BdG}}(-\mathbf{k})\quad.
    \label{eq:bdgcondition}
\end{align}

In this formalism the $U(1)$ symmetry is not automatic, it is generated by $\tau^z$. We can add a $U(1)$ breaking term such as $(1-\cos k_z)\tau^x\sigma^y$ to gap the Weyl node at $\mathbf{k}_2$, leading to the modified Hamiltonian
\begin{align}
    h_\text{single Weyl}^{\mathrm{BdG}}(\mathbf{k})= &\frac{1}{2}\,[\sin k_x\sigma^x+\sin k_y \sigma^z +m(\mathbf{k})\sigma^y\nonumber\\
    &+\sin k_z\tau^z\sigma^y+(1-\cos k_z)\tau^x\sigma^y]\,.
\end{align}
This Hamiltonian has a single Weyl node remaining at $\textbf{k}_1 = 0$, and no other gapless modes{, when $|\Delta|>2$}.

It turns out $h_\text{single Weyl}^{\mathrm{BdG}}(\mathbf{k})$ commutes with a symmetry generator $S_\text{chiral}(\mathbf{k})$ given by
\begin{align}\label{eqnsymmetry1}
    S_\text{chiral}(\mathbf{k})=\frac{1}{2}\left[(1+\cos k_z)\tau^z+\sin k_z \tau^x\right]\quad.
\end{align}
At $\mathbf{k}_1$ the symmetry reduces to $S_\text{chiral}(\mathbf{k}_1)=\tau^z$ which is just the original $U(1)$ charge operator for the corresponding single-particle modes. Therefore, it gives an exact chiral symmetry in this model.

We can see by inspection that this symmetry prevents all mass terms, but allows for terms such as $\sigma^y$ and $(1+\cos k_z)\tau^z+\sin k_z\tau^x$ which only shifts the Weyl node in momentum space. Thus, the anomaly indeed stabilizes the low energy theory. One interesting caveat is that at $\mathbf{k}_2$, $S_\text{chiral}(\mathbf{k}_2)=0$, which is what allowed us to gap the second Weyl node there. We could eventually symmetrically move the remaining Weyl node to this point as well and then completely gap the system.

We can write the associated charge operator
\begin{align}\label{eqnchiralgenerator}
    \hat{Q}_\text{chiral}=\sum_\mathbf{k} d_{\mathbf{k}}^\dag \,S_\text{chiral}(\mathbf{k})\,d_{\mathbf{k}}\quad,
\end{align}
via a Fourier transform in real space as
\begin{align}
    \hat{Q}_\text{chiral}=\frac{1}{2}\sum_{\mathbf{\mathbf{r}},s} \bigg[c_{\mathbf{r},s}^\dag c_{\mathbf{r},s} +c_{{\mathbf{r}+\hat{z}},s}^\dag c_{\mathbf{r},s}-ic_{{\mathbf{r}+\hat{z}},s}^\dag c_{\mathbf{r},s}^\dag\bigg]+h.c.
\end{align}
We see that, like the Ginsparg-Wilson symmetry, this charge operator involves nearest neighbor couplings, and is thus not-on-site.

{

Moreover, we see that the BdG generator obeys $S_\text{chiral}(\mathbf{k})^2 = \cos^2(k_z/2) \mathds{1}$. Curiously, as a Hamiltonian itself, it describes decoupled wires along the $z$-axis of massless Majorana fermions.\footnote{A \emph{non-Hermitian} symmetry generator of a single Weyl fermion was proposed in \cite{catterall2025symmetriesanomalieshamiltonianstaggered}, by interpolating between the identity and a translation symmetry.}  Thus, it has a continuous spectrum, and is therefore a non-compact symmetry, generating an $\bR$ action on the full Hilbert space. 

In fact, a non-quantized spectrum is necessary to evade the no-go theorem of Fidkowski and Xu \cite{Fidkowski_2023}, a generalization of Nielsen-Ninomiya. Their theorem states that a chiral $U(1)$ symmetry with an exponentially-local charge operator cannot act non-trivially on single Weyl fermion in the IR\footnote{This was recently emphasized and explored in \cite{singh2025ginspargwilsonhamiltoniansimprovedchiral}.}. We state and prove a closely related no-go theorem below.}

\section{A No-Go Theorem for $U(1)$ Symmetries}

{ 

We give a no-go theorem along the lines of Fidkowski and Xu \cite{Fidkowski_2023} in the context of free-fermion symmetries in arbitrary dimensions.

Let $\hat Q_\text{chiral}$ be a generator defining a $U(1)$ symmetry (i.e. with quantized charges), and assume that $\hat Q_\text{chiral}$ is a free-fermion symmetry, and is generated by a translation-invariant single-particle Hamiltonian $S_\text{chiral}$ with exponential decay in real space.
We will show \textit{$\hat Q_\text{chiral}$ admits a symmetric, gapped, exponentially-local Hamiltonian which is non-degenerate on a torus}.

First we observe that $\hat Q_\text{chiral}$ itself may be viewed as a Hamiltonian with chiral symmetry. Indeed, it is Hermitian and satisfies $[\hat Q_\text{chiral},\hat Q_\text{chiral}]=0$. By our assumption of the $U(1)$ symmetry, $\hat Q_\text{chiral}$ has an integer spectrum, so $S_\text{chiral}$ does as well. If $S_\text{chiral}$ has no zero eigenvalue, then the ground state of $\hat Q_\text{chiral}$ is the state with all negative eigenvalues of $S_\text{chiral}$ filled and all positive eigenvalues of $S_\text{chiral}$ empty, with an energy gap to the first positive eigenvalue. This is a band insulator and is non-degenerate on a torus and in this case is what we wanted to show.

Suppose on the other hand that $S_\text{chiral}$ has a zero eigenvalue. Then $\hat Q_\text{chiral}$ has an extensively degenerate flat band in its ground state, with a gap to its further excited states, and we cannot immediately conclude as we did above. Consider the single-particle operator $P(\textbf{k})$ which in the eigenbasis of $S_\text{chiral}(\textbf{k})$ acts as $1$ on the zero eigenvectors of $S_\text{chiral}(\textbf{k})$ and as $0$ on the other eigenvectors. The single-particle Hamiltonian $h(\textbf{k})=S_\text{chiral}(\textbf{k}) + P(\textbf{k})$ has no zero energy band and commutes with $S_\text{chiral}(\textbf{k})$. In real space, it is at least exponentially localized since $S_\text{chiral}$ has a uniform band gap of 1 (see Appendix \ref{appspectralprojector}). Therefore, it defines a local Hamiltonian $\hat H$ with $\hat Q_\text{chiral}$ symmetry. Since we have removed the zero energy band, the ground state of $\hat H$ is a gapped band insulator as above, finishing the argument.

By standard anomaly-matching lore, having such a ``trivial'' symmetric Hamiltonian means that $\hat Q_\text{chiral}$ is free of 't Hooft anomalies. It therefore cannot realize the chiral symmetry of a single Weyl fermion, by anomaly matching.
}

\section{Symmetry-protected double Weyl fermion in 3+1D}\label{secSU2}

The symmetry \eqref{eqnsymmetry1}, which has continuous spectrum, may be naturally separated into two generators with quantized spectrum:
\[
 S_0(\textbf{k}) = \tau^z \qquad \hat{Q}_0=\sum_\mathbf{k} d_{\mathbf{k}}^\dag S_0(\mathbf{k})d_{\mathbf{k}} \\
 S_1(\textbf{k}) = \cos k_z \tau^z + \sin k_z \tau^x \qquad \hat{Q}_1=\sum_\mathbf{k} d_{\mathbf{k}}^\dag S_1(\mathbf{k})d_{\mathbf{k}}.
\]
The first is the usual $U(1)$ symmetry, while the second is composed of Kitaev Majorana chains \cite{kitaev2001unpaired} along $z$-axis wires. These generators do not commute, instead they generate an infinite-dimensional Lie algebra known as the Onsager algebra, introduced in \cite{onsager1944crystal}. This algebra has recently appeared in the study of the 1+1D chiral anomaly on the lattice \cite{Vernier_2019,jones2024pivotingchiralclockfamily,chatterjee2025quantized}, and offers another route to defining exact symmetries on the lattice giving anomalous symmetries in the IR.

We can actually write a Hamiltonian that has this symmetry and \emph{two} Weyl nodes, which in the BdG formalism above (using the $d_\mathbf{k}^\dag$ basis) is
\begin{align}
    h_\text{double Weyl}^{\mathrm{BdG}}(\mathbf{k})= &\mathds{1}_\tau\otimes \frac{1}{2}\bigg[\sin k_x\sigma^x+\sin k_y \sigma^z\nonumber\\
    &+\left[\cos k_z-\cos K+m(\mathbf{k})\right]\sigma^y\bigg]\quad,
    \label{eq:2weylsym}
\end{align}
where the identity in the $\tau$ basis ensures it has both $U(1)$ symmetries, $K$ is a parameter, and $m(\textbf{k})$ is the same as in \eqref{eqnh2}.
{When $|\cos K-\Delta|>1$,} this model is a magnetic Weyl semimetal model {with} two Weyl nodes { of opposite chirality} at $\textbf{k} = \pm \mathbf{K}$ where $\mathbf{K}=(0,0,K)$.

Let's linearize around the Weyl nodes. We get
\[h^\text{BdG}_l = \mathds{1}_\tau \otimes \frac{1}{2}(k_x \sigma^x + k_y \sigma^z -\sin K\ k_z \sigma^y).\]
This shows that for $K \neq 0, \pi$, the two Weyl nodes have an opposite handedness. To figure out the effect of $\hat S_1(\textbf{k})$ at low energy, we can also linearize it, and obtain
\[ S_{1,K}(\textbf{k}) = \cos K\ \tau^z + \sin K\ \tau^x.\]
The important feature for $K \neq 0,\pi$ is that the second term is non-zero.

Thus, together with $S_0 = \tau^z$, these generate an $su(2)$ algebra acting on the low energy theory. Note that $\tau^x$ acts by exchanging particles at $\textbf{K}$ with holes at $-\textbf{K}$. Thus, it is convenient to apply a charge conjugation the right-handed Weyl fermion, to give a low energy theory in terms of two left-handed Weyl fermions, now with \emph{opposite} charge w.r.t. $\tau^z$ rotations. $\tau^x$ rotations meanwhile act by a flavor rotation exchanging the two Weyl fermions. Thus, our symmetry generators $\hat Q_0$, $\hat Q_1$ correspond to two $su(2)$ generators in the flavor symmetry of the low energy, at an angle of $K$. For $K \neq 0, \pi$, they thus generate the whole chiral symmetry.

We can demonstrate that this symmetry protects the gapless Weyl points. To do so, we must break translation symmetry, since otherwise $z$-axis translations also act as a discrete axial symmetry and help to stabilize the Weyl nodes \cite{PhysRevLett.124.096603,PhysRevResearch.3.043067}. To analyze translation-symmetry breaking, we consider an extended basis $e_{\mathbf{k}}\equiv(c_{\mathbf{k}-\mathbf{K}},c^\dag_{-\mathbf{k}+\mathbf{K}},c_{\mathbf{k}+\mathbf{K}},c^\dag_{-\mathbf{k}-\mathbf{K}})^T$ (we suppress the spin component). Hamiltonians in this basis may couple states at $\mathbf{k}$ with $\mathbf{k}+2\mathbf{K}$ but are automatically $\hat Q_0$ preserving. In this basis the symmetry action of $\hat Q_1$ becomes
\begin{align}
    U(1)_{\hat Q_1}:\delta e_{\mathbf{k}} = &i(\cos k_z \cos K\ \tau^z +\sin k_z \sin K\ \,\eta^z\tau^z\nonumber\\
    +&\sin k_z\cos K\ \tau^x-\cos k_z\sin K\ \,\eta^z\tau^x)e_{\mathbf{k}},
\end{align}
which prohibits all mass terms except $m_j(\mathbf{k})\eta^z\sigma^j$. However, these terms always commute with at least one term in the original Hamiltonian, so the result is a shift in the gapless modes rather than a gap. At a large enough pertubation, we can move the modes until $K = 0$ or $\pi$, where the symmetry generators are aligned and no longer generate the whole $SU(2)$ symmetry. At these special points, we will be able to open a symmetric gap.

\section{Discussion}

In this work, we have introduced several Hamiltonian models with new, not-on-site symmetries giving rise to anomalous symmetries acting on the low energy fermions. We have realized the chiral anomaly of a single charged Weyl fermion in 3+1D, the $SU(2)$ anomaly of a doublet of left-handed Weyl fermions in 3+1D, and the $U(1) \rtimes T$ parity anomaly of a single Dirac fermion in 2+1D.

The chiral symmetry in 3+1D has a nearest-neighbor charge operator, but is non-quantized, as it must be by the no-go theorem of \cite{Fidkowski_2023} and our arguments above. The two $SU(2)$ generators we constructed are quantized (and one is on-site), but do not satisfy the expected $SU(2)$ Lie algebra, instead generating an infinite-dimensional Onsager algebra. This is consistent with the anomaly since either $U(1)$ symmetry on its own is anomaly-free. In this way, it is structurally similar to the $U(1)_V \times U(1)_A$ anomaly realized in 1+1D in \cite{jones2024pivotingchiralclockfamily,chatterjee2025quantized}, where both generators are quantized and one is on-site, but they don't commute.

To include time reversal symmetry in these systems, we had to relax our charge density operators to be almost-local operators, with faster-than-polynomial decaying tails. We are unsure if this is a necessary further weaking of on-siteness, but it is convenient.

In each example, we are able to show a no-go theorem that shows our construction is nearly as good as possible. The method for proving these no-go theorems seems very general for studying anomalous $U(1)$ symmetries. If we have a $U(1)$ symmetry generator $\hat Q$, which is either anomalous or shares an anomaly with other symmetries commuting with it. Then we can regard $\hat Q$ itself as a Hamiltonian with these anomalous symmetries, including $\hat Q$. Therefore, $\hat Q$ must have a non-trivial ground state.

There is a fun example which avoids this no-go argument in 2+1D. Let us take a 2d square lattice of spin-$\frac12$ degrees of freedom, and take $\hat Q$ to be the toric code Hamiltonian \cite{Kitaev_2003}, which has an integer spectrum. It also commutes with time reversal given by complex conjugation, and together, these two symmetries generate a bosonic $U(1) \rtimes T$ parity anomaly \cite{Tantivasadakarn_2023}. The ground state of the toric code Hamiltonian is indeed non-trivial { (it is degenerate on a torus)}, and sufficient to match this parity anomaly.

Finally, we would eventually like to make lattice models of chiral gauge theories. Although we have produced models with chiral symmetries, the not-on-siteness of the symmetries makes them difficult to gauge. In fact one expects on general grounds that because of the nonvanishing 't Hooft anomalies, we should not be able to gauge these symmetries. For interesting anomaly cancellation, such as in the Standard Model, one needs low energy fermions of different charges. It is not clear models of the kind we studied can produce such charge assigments. It would be very interesting if in some more complicated models with not-on-site chiral symmetries having vanishing 't Hooft anomalies, if we can nonetheless find some way to gauge them, as we done for discrete symmetries in 1+1D in \cite{Seifnashri_2024}.

\textit{Acknowledgements:} We are grateful for conversations with David Kaplan, Shu-Heng Shao, and Cenke Xu. LG acknowledges support from the Walter Burke Institute for Theoretical Physics at
Caltech and the Caltech Institute for Quantum Information and Matter. RT acknowledges support from the Mani L. Bhaumik Presidential Term Chair in Theoretical Physics at UCLA. 

\bibliography{arxiv}

\appendix
{

\section{Fermion Doubling and Anomalies}\label{appfermiondoubling}

We can also reason about fermion doubling from the point of view of 't Hooft anomaly matching \cite{luscher2001chiral}. In particular, a system with an on-site global symmetry, meaning one which does not mix degrees of freedom at separate spacetime points, is free of 't Hooft anomalies in the UV, since a 't Hooft anomaly is an obstruction to gauging and there is a canonical method to gauge on-site symmetries. By anomaly matching, whatever theory arises from this lattice model in the IR must also be free of 't Hooft anomalies.

For an IR composed of $N_L$ left-handed Weyl fermions of charges $q_i^L$ and $N_R$ right-handed Weyl fermions of charge $q_i^R$, the anomaly-vanishing equations are
\[\sum_{i=1}^{N_L} q_i^L = \sum_{i=1}^{N_R} q_i^R \qquad \sum_{i=1}^{N_L} (q_i^L)^3 = \sum_{i=1}^{N_R} (q_i^R)^3.\]
Trivial solutions are those with $N_L = N_R$ and $q_i^L = q_i^R$ where fermions come in non-chiral pairs. For chiral gauge theories like the Standard Model, we want the IR to realize a non-trivial solution to these equations.

However, in a free-fermion (quadratic) Hamiltonian model, fermions of different charge never interact, and so the anomaly-vanishing equation has to hold for each $U(1)$ charge separately, yielding only trivial solutions with fermion doubling.

\section{Cubic-symmetric single Weyl fermion model}
\label{app:cubicsym}

In this appendix we will present a model of a symmetry-protected single Weyl fermion that is fully cubic-symmetric, as opposed to the the model in Section~\ref{sec:symprot1weyl}. Here, we will take the BdG basis $d_\mathbf{k}^\dag=(c_{\mathbf{k}}^\dag,[i\sigma^yc_{-\mathbf{k}}]^T )$, which imposes the particle-hole redundancy
\begin{align}
    \tau^y\sigma^y h^{\mathrm{BdG}}(\mathbf{k})^T \tau^y\sigma^y=-h^{\mathrm{BdG}}(-\mathbf{k})\quad,
\end{align}
as opposed to Eq.~\ref{eq:bdgcondition}.

Consider the Bloch Hamiltonian in the BdG formalism given by
\begin{align}
    h^{\rm BdG}(\mathbf{k})=\sin k_x&\tau^z\sigma^x+\sin k_y\tau^z\sigma^y\nonumber\\
    &+\sin k_z\tau^z\sigma^z+M(\mathbf{k})\tau^x\quad,
    \label{eq:hbdgcubic}
\end{align}
where $M(\mathbf{k})=3-\cos k_x-\cos k_y-\cos k_z$. This model possesses a single massless Weyl fermion at $\mathbf{k}=0$, and retains the cubic symmetry of the lattice, with the $\frac{\pi}{2}$ rotations given by $i\sigma^{x,y,z}$ as well as time-reversal symmetric. A similar model was given in the original Nielsen-Ninomiya paper~\cite{NIELSEN1981173}.

We can construct a chiral symmetry generator $S_{\rm chiral}(\mathbf{k})$ that protects this massless Weyl fermion, with $S_{\rm chiral}(\mathbf{k})$ given by
\begin{align}
    S_{\rm chiral}(\mathbf{k})=&\frac{1}{8}\big[(1+\cos k_x)(1+\cos k_y)(1+\cos k_z)\tau^z\nonumber\\
    &+\sin k_x (1+\cos k_y)(1+\cos k_z)\tau^x\sigma^x\nonumber\\
    &+ (1+\cos k_x)\sin k_y(1+\cos k_z)\tau^x\sigma^y\\
    &+ (1+\cos k_x)(1+\cos k_y)\sin k_z \tau^x\sigma^z\big],\nonumber
\end{align}
where $[S_{\rm chiral}(\mathbf{k}),h^{\rm BdG}(\mathbf{k})]=0$. Observe that
\begin{align}
S_{\rm chiral}(\mathbf{k})=
\begin{cases}
    \tau^z &{\rm when}\,\,\mathbf{k}=\mathbf{0}\\
    0 &{\rm when}\,\,k_x=\pi\\
    0 &{\rm when}\,\,k_y=\pi\\
    0 &{\rm when}\,\,k_z=\pi
\end{cases}\quad,
\end{align}
i.e., $S_{\rm chiral}(\mathbf{k})$ acts like the usual $U(1)$ symmetry $\tau^z$ at $\mathbf{k}=0$, where the massless Weyl fermion is, but is $0$ at the boundary of the Brillouin zone. This symmetry is still unquantized, i.e. the charge generator is not quantized. The physical interpretation of such a symmetry, when Fourier transformed back to real space, is not as simple the bundle of critical Kitaev chains in the example given in Eq.~\ref{eqnsymmetry1} of the main text. In this case, the real space Hamiltonian would involve fermion hoppings of range three, whose ground state corresponds to a Bogoliubov Fermi surface state.

The Hamiltonian in Eq.~\ref{eq:hbdgcubic} with the chiral symmetry, generated by $S_{\rm chiral}(\mathbf{k})$, still possesses terms that can eventually (non-perturbatively) gap the single massless Weyl fermion since the system does not possess a quantized UV anomaly~\footnote{ A more detailed handling of quantized versus unquantized anomalies can be found in Ref.~\cite{gioia2025nogotheoremsingletimereversal}.}. $S_{\rm chiral}(\mathbf{k})$ is itself such a term -- upon adding it with a magnitude $\delta>0$ to Eq.~\ref{eq:hbdgcubic}, the Weyl fermion at the origin transitions into a Fermi surface with the low-energy linearization around $\mathbf{k}=\mathbf{0}$ being
\begin{align}
    h^{\rm BdG}(\mathbf{k})\approx k_x&\tau^z\sigma^x+k_y\tau^z\sigma^y+ k_z\tau^z\sigma^z+\delta\tau^z\quad,
\end{align}
when $\delta$ is small. When the magnitude $\delta$ is increased, this Fermi surface expands and then eventually gaps at the boundary of the Brillouin zone.

\section{Time Reversal Symmetric Single Weyl Fermion in 3+1D}

So far, we have considered time-reversal breaking models which are ultralocal in space, having bounded-range hoppings, with symmetries satisfying the same. We can also construct time-reversal invariant models, but this will come at the cost of making the models and their symmetries non-ultralocal. We do not have a proof that it is necessary to do this, and in the future it would be interesting to explore whether it is necessary is not.

As long as $S(\textbf{k})$ is a smooth function of the momentum, then the charge density in real space will be a sum of terms with faster-than-polynomial decay. Such ``almost-local'' operators share many properties with local operators. In particular, time evolution under almost-local operators is locality-preserving, in the sense of having a Lieb-Robinson bound \cite{kapustin2022local}, mapping almost-local operators to almost-local operators.

Allowing these almost-local operators will allow us to employ bump functions and partitions of unity in momentum space and greatly simplifies the constructions.

To construct a time-reversal invariant model with a single protected Weyl fermion, we begin with a model on a cubic lattice with eight Weyl nodes. We use the BdG formalism with the basis $d^\dag_\mathbf{k}\equiv(c^\dag_{\mathbf{k}\uparrow},c^\dag_{\mathbf{k}\downarrow},c_{-\mathbf{k}\uparrow},c_{-\mathbf{k}\downarrow})$ used above, giving the Hamiltonian
\begin{align}
    h^{\mathrm{BdG}}_8(\mathbf{k})=\frac{1}{2}\left[\sin k_x\sigma^x+\sin k_y \sigma^z +\sin k_z\tau^z\sigma^y\right].
\end{align}
This model has Weyl nodes at all eight time-reversal-invariant-momentum (TRIM) points of the Brillouin zone, as well as a time-reversal symmetry $\Theta = i\sigma^y \mathcal{K}$, where $\mathcal{K}$ is complex conjugation, satisfying $\Theta^2=-1$.

We will now add a $U(1)$ symmetry-breaking term that will gap out all Weyl nodes except the one at $\mathbf{k=0}$. In order to facilitate our discussion, let us first define a bump function $B(k)$ given by
\begin{align}
    B(\mathbf{k},w)=
    \begin{cases}
    e^{\frac{|\mathbf{k}|^2}{|\mathbf{k}|^2-w^2}} & \mathrm{for}\,\,|\mathbf{k}|<w\\
    0 & \mathrm{for}\,\,|\mathbf{k}|\geq w
    \end{cases}
\end{align}
where $w>0$ determines the width of the bump. This function is smooth but non-analytic. We add a $U(1)$ breaking term such that the total Hamiltonian is now given by
\begin{align}
    h^{\mathrm{BdG}}_{\text{TRS Weyl}}(\mathbf{k})=&  \nonumber \\ h_8^\text{BdG}(\textbf{k})+  \sum_{j\in\{x,y,z\}}&B\left(k_j-\pi,\frac{\pi}{2}\right)(1-\cos k_j)\tau^y\sigma^y\quad,
\end{align}
which gaps all Weyl nodes except the one at $\mathbf{k}=0$. By inspection, this preserves the time-reversal symmetry $\Theta$. It also has an almost-local chiral symmetry generator
\begin{align}\label{eqnbumpchiral}
    S_\text{chiral}(\mathbf{k})=B\left(\mathbf{k},\frac{\pi}{2}\right)\tau^z
    \quad.
\end{align}
As previously, this chiral symmetry is not quantized, as it must be by \cite{Fidkowski_2023} and our arguments in the previous section. We could also choose a step function instead of a bump function for this symmetry, and get a quantized chiral symmetry, but in real space it would not be almost-local, with the charge density having algebraic decay. This also avoids \cite{Fidkowski_2023} because such an operator does not generate a locality preserving unitary evolution.

\section{Parity anomaly of a Single Dirac fermion with time-reversal symmetry in 2+1D}

We start with a 2+1D time-reversal invariant Dirac fermion model on a square lattice with four Dirac nodes given by the BdG Hamiltonian with basis $d^\dag_\mathbf{k}\equiv(c^\dag_{\mathbf{k}\uparrow},c^\dag_{\mathbf{k}\downarrow},c_{-\mathbf{k}\uparrow},c_{-\mathbf{k}\downarrow})$
\begin{align}
    h_{4}^\text{BdG}(\mathbf{k})= \mathds{1}_\tau \otimes\frac{1}{2}(\sin k_x\sigma^x+\sin k_y\sigma^y)\quad.
\end{align}
This model has Dirac nodes are at all four TRIM points of the Brillouin zone, and a time-reversal symmetry $\Theta = i\sigma^y \mathcal{K}$ with $\Theta^2=-1$. We will now add a $U(1)$ symmetry-breaking term that will gap out all Dirac nodes except the one at $\mathbf{k=0}$:
\begin{align}
    h^{\mathrm{BdG}}_{\text{single Dirac}}(\mathbf{k})= &  \\  h_4^\text{BdG}(\textbf{k}) + &\sum_{j\in x,y}B\left(k_j-\pi,\frac{\pi}{2}\right)(1-\cos k_j)\tau^y\sigma^y,\nonumber 
\end{align}
This gaps all Dirac nodes except the one at $\mathbf{k}=0$.
This model commutes with an almost-local symmetry generator of the same form as \eqref{eqnbumpchiral}:
\begin{align}
    S(\mathbf{k})=B\left(\mathbf{k},\frac{\pi}{2}\right)\tau^z,
\end{align}
which commutes with time-reversal $\Theta$ and again has continuous spectrum. Since at $\mathbf{k} = 0$, it acts as $\tau^z$, it gives rise to the $U(1)$ symmetry of the single Dirac fermion in the IR, which together with $\Theta$ protects this Dirac fermion from gaining a mass by the parity anomaly. Note that, analogous to the previous examples, by a large enough perturbation, we can push the Dirac cone into the region where $S(\textbf{k}) = 0$ and eventually gap out the system.

As with the 3+1D chiral symmetry, this particle number symmetry of this type---acting linearly on fermions and commuting with a time-reversal action---must be non-quantized. This follows the same argument as in 3+1D. Otherwise, we could consider the $U(1)$ symmetry generator itself as a $U(1) \rtimes T$ symmetric Hamiltonian. For these two band models, it would describe a band insulator with a parity anomaly, which is impossible.

\section{Locality of the Spectral Projector}\label{appspectralprojector}

Suppose $h(\textbf{k})$ describes a single-particle Hamiltonian which is exponentially decaying in real space, with a zero-energy band and all other bands occur outside the energy window $(-1,1)$. We consider the ``Riesz projector''
\[P(\textbf{k}) = \frac{1}{2\pi i}\oint_C \frac{dz}{z-h(\textbf{k})},\]
where $C$ is a positively-oriented circle of radius $<1$ centered around the origin. Going to the eigenbasis of $h(\textbf{k})$ and evaluating by residues, this gives the projector we discussed in the main text (with $h(\textbf{k}) = S_\text{chiral}(\textbf{k})$).

Consider analytically continuing this expression to complex $\textbf{k}$. Since $h(\textbf{k})$ corresponds to an exponentially decaying Hamiltonian in real space, it is analytic in some strip $|{\rm Im}(k_i)| < \epsilon$. Also in this strip, the zeros of $z-h(\textbf{k})$ move continuously in the $z$ plane, so we can choose some sub-strip $|{\rm Im}(k_i)| <\epsilon_1 \le \epsilon$ where no zero crosses our small circle $C$. Therefore, in that strip $P(\textbf{k})$ will be analytic and so $P(\textbf{k})$ is exponentially-decaying in real space.

}

\end{document}